\documentclass[%
reprint,
amsmath,amssymb,
aps,
]{revtex4-2}

\usepackage{graphicx}% Include figure files
\usepackage{dcolumn}% Align table columns on decimal point
\usepackage{bm}% bold math
\begin{document}

\preprint{APS/123-QED}

\title{Example Exact Solutions of the Time-Independent 
{G}ross-{P}itaevskii 
and {S}chr{\"o}dinger 
Equations}

\title{Example Exact Solutions of the Time-independent Gross-Pitaevskii and Schr{\"o}dinger Equations} % Force line breaks with \\
\author{Bhavika Bhalgamiya \textsuperscript{1}}
 \altaffiliation{Department of Physics and astronomy, Mississippi State University, Mississippi State, Mississippi 39762-5167, USA}%Lines break automatically or can be forced with \\
\author{M. A. Novotny \textsuperscript{1}}%
 \email{man40@msstate.edu}
\affiliation{$^1$Department  of Physics and Astronomy 
and HPC$^2$ Center for Computational Sciences, 
Mississippi State University, Mississippi State, Mississippi 39762-5167, USA
}%%
%\affiliation{$^2$Faculty of Mathematics and Physics, %Charles University in Prague, Ke Karlovu 5, CZ-121 16 Praha %2, Czech Republic}

\date{\today}% It is always \today, today,

\begin{abstract}
A prescription is given to obtain some exact results for certain 
external potentials $V\left({\vec r}\right)$ of the time-independent 
Gross-Pitaevskii and Schr{\"o}dinger equations.  
The study motivation is the ability to program 
$V\left({\vec r}\right)$ experimentally in Bose-Einstein condensates.  
Rather than derive wavefunctions that are solutions for a given $V\left({\vec r}\right)$, we 
ask which $V\left({\vec r}\right)$ will have a given 
pdf (probability density function) $P\left({\vec r}\right)$.  
Several examples in 1D and 2D are presented for well-known pdfs and for 
the hydrogen atom in momentum space.  
\end{abstract}

\maketitle 

Almost a century ago, Bose \cite{bose1924plancks} and 
Einstein \cite{einstein2005quantentheorie} introduced what is sometimes 
called the fifth state of matter, the Bose-Einstein Condensate (BEC). 
It was not until 1995 
\cite{Petrich1995,Anderson1995,davis1995}
that a BEC was created in a laboratory by 
using gases of ultra-cold atoms of $^{87}$Rb and $^{23}$Na. The 2001 Nobel 
prize in physics was awarded to these researchers for the experimental demonstration 
of a BEC \cite{cornell2002nobel,ketterle2002nobel}. 
The development of atomtronics attempts to use a BEC condensate for 
quantum sensing, quantum computing, and quantum information 
science \cite{amico2021roadmap,pepino2021advances}. 
The project reported here is motivated by the recent introduction of a stand-alone 
device by the company ColdQuanta \cite{ColdQuanta} where 
an atomic gas BEC can be created, 
and furthermore the BEC is in an external potential $V({\vec r})$ 
which can be \lq painted' to have particular values in 2D (2 dimensions). 
\lq Painted' potentials for atomic BEC systems have also been 
reported by other groups \cite{Henderson2009}.  

The underlying equation for a BEC in a gas has been 
rigorously shown \cite{ERDOS2007}, with 
reasonable assumptions, to be the Gross-Pitaevskii equation 
(GPeq) \cite{gross1961structure,pitaevskii1961vortex}.
The GPeq is one type of a non-linear Schr{\"o}dinger equation \cite{meyer2014quantum}. 
If the gas used for the BEC has $N$ atoms of mass $m$ and a scattering length $a_s$, 
the nonlinear term in the GPeq is $g = 4\pi\hbar^2a_s/m$.  
Here $\hbar$ is Plank's constant divided by $2\pi$. 
Note that in an atomic BEC $g$ can sometimes be changed 
experimentally by orders of magnitude and even 
change its sign \cite{Donley2001,Altin2010}.  
The time-independent GPeq is expressed as \cite{davis2001dynamics}
\begin{equation}
\label{Eq:GPeq}
\mu{\widetilde\psi}({\vec r}) = 
\left(-\frac{\hbar^2}{2m}\bigtriangledown^2+
{\widetilde V}({\vec r})\right){\widetilde\psi}({\vec r})
\end{equation}
with $\mu$ chemical potential. 
Here ${\widetilde\psi}({\vec r})$ is the wave function of the BEC, and 
\begin{equation}
\label{Eq:VandV}
{\widetilde V}({\vec r}) = V({\vec r})+
gN\left|{\widetilde\psi}({\vec r})\right|^2 
\>.
\end{equation}

We also detail a prescription to find solutions of the 
time-independent Schr{\"o}dinger equation (TISE), namely
\begin{equation}
\label{Eq:TISE}
-\frac{\hbar^2}{2m} \bigtriangledown^2\psi({\vec r}) 
+ \left[V({\vec r})-E\right]\psi({\vec r}) = 0
\end{equation}
where for a single mass $m$ particle the wavefunction is 
$\psi({\vec r})$ with energy $E$.  
Note the differences in the physical interpretation between Eq.~(\ref{Eq:GPeq}) 
for ${\widetilde\psi}$ and the 
TISE of Eq.~(\ref{Eq:TISE}) for $\psi$.  
Nevertheless, mathematically Eq.~(\ref{Eq:GPeq}) goes to 
Eq.~(\ref{Eq:TISE}) with the replacements 
${\widetilde\psi}\rightarrow\psi$, $\mu\rightarrow E$, and $gN=0$ so 
${\widetilde V}\rightarrow V$.  

Since the first encounter with the TISE a student learns the traditional way 
to solve Eq.~(\ref{Eq:TISE}).  
By the term 'traditional approach'  we mean:
given a $V({\vec r})$ solve for $\psi({\vec r})$ and $E$.  Then find the 
pdf using $P({\vec r})=\left|\psi({\vec r})\right|^2$.  
Several interesting applications of this approach are explained in 
standard textbooks, including \cite{griffiths2018introduction,mcintyre2012quantum},
for real space as well as momentum space. 
As opposed to the traditional method, we propose an alternative prescription.   
Namely, given a pdf $P({\vec r})$ find which potential $V({\vec r})$ and energy $E$ 
in the 
TISE is needed with $\psi({\vec r})=\sqrt{P({\vec r})}$ to give the desired pdf.  
Our prescription works only in some cases, requiring certain mathematical 
properties for $P({\vec r})$.  Nevertheless, our prescription also works 
in certain cases to obtain solutions of the GPeq for general $g N$, as well as 
for some other 
nonlinear Schr{\"o}dinger equations.  

We first outline the prescription for the TISE of Eq.~(\ref{Eq:TISE}).  
In order to form the 3D potential $V({\vec r})$ 
for the given 3D pdf $P({\vec r})$, we define an 
analytic function $f({\vec r})$.  
We assume we can choose the wavefunction to be real, thereby limiting which 
problems we can solve.  We assume  
\begin{equation}
\label{Eq:define:f}
\psi({\vec r}) \> = \> \sqrt{A} \> \exp\left(- f({\vec r})/2\right)
\end{equation}
so $P({\vec r})=A \exp\left(-f({\vec r})\right)$, with $A$ for the normalization 
$\int P({\vec r}) \>d{\vec r}\>=\>1$.  
Using first and second-order derivatives of the function $f({\vec r})$, 
we can solve both the time independent Schr{\"o}dinger and Gross-Pietevskii equations.  

For the TISE, by substituting Eq.~(\ref{Eq:define:f}) into Eq.~(\ref{Eq:TISE}) 
we obtain the equation 
\begin{equation}
\label{Eq:TISE:f}
V({\vec r})-E =  
- \frac{\hbar^2}{4m}
\nabla^2 f({\vec r})
+\frac{\hbar^2}{8m}\left[\left({\vec\nabla}f({\vec r})\right)\cdot
\left({\vec\nabla}f({\vec r})\right)\right]
\>.
\end{equation}
For the time-independent GPeq we also use the {\it ansatz\/} 
${\widetilde\psi}({\vec r})=\sqrt{A} \exp[-f({\vec r})/2]$ and obtain 
the equation 
\begin{equation}
\label{Eq:GPeq:f}
\widetilde{V}({\vec r})-\mu= 
-\frac{\hbar^2}{4m}\nabla^2 f({\vec r})
+
\frac{\hbar^2}{8m}\left[
\left({\vec\nabla}f\right)\cdot\left({\vec\nabla}f\right)
\right]
\>.
\end{equation}

\begin{figure}[htp]
    \centering
    \includegraphics[width=8.6 cm]{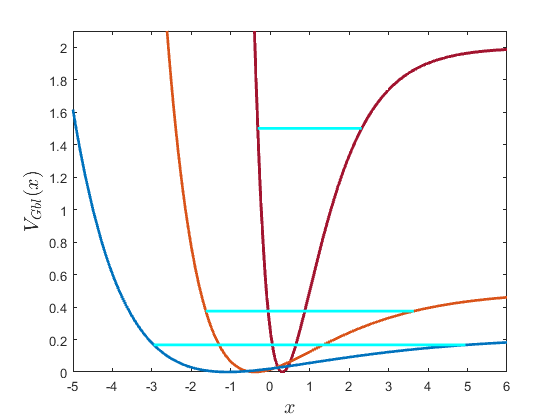}
    \caption{Derived potential $V_{\rm Gbl}(x)$ for the 1D Gumbel distribution 
    for the TISE, with units $\frac{\hbar^2}{4m}= 1$, and $x_0 = 1$ with $\beta = 1$ 
    (maroon) $\beta=2$ (orange) and $\beta= 3$ (blue).  The horizontal cyan lines 
    are the associated groundstate energies $E$ for each $\beta$.}  
    \label{Fig:Gumbel:TISE}
\end{figure}

We expand on the prescription applied to the 1D Gumbel distribution for both 
Eq.~(\ref{Eq:TISE:f}) and Eq.~(\ref{Eq:GPeq:f}).  
The Gumbel pdf is 
\begin{equation}
\label{Eq:Gumbel:pdf}
P_{\rm Gbl}(x) = 
\frac{1}{\beta}
\exp{\left[-\left(\frac{x-x_0}{\beta}+
\exp{\left[-\frac{x-x_0}{\beta}\right]}\right)\right]} 
\end{equation}
where $x_0$ is the mode of the pdf, and $\beta$ is a scale parameter. 
From  Eq.~(\ref{Eq:Gumbel:pdf}) and in 1D Eq.~(\ref{Eq:define:f}), 
the function $f_{\rm Gbl}(x)$ associated with the distribution $P_{\rm Gbl}(x)$ 
\begin{equation}
\label{Eq:Gumbel:f}
f_{\rm Gbl}(x) = \frac{x-x_0}{\beta}+\exp\left[-\frac{x-x_0}{\beta}\right]+\ln{\beta}
\end{equation}
with $A=1$.  
The potential $V_{\rm Gbl}(x)$ for the 1D Gumbel pdf 
is obtained by substituting the first and second-order derivatives 
of Eq.~(\ref{Eq:Gumbel:f}) into Eq.~(\ref{Eq:TISE:f}) in 1D.  This gives,
choosing the zero of energy as the minimum of $V_{\rm Gbl}(x)$, as 
\begin{equation}
\label{Eq:Gumbel:V}
V_{\rm Gbl}(x) = 
\frac{\hbar^2}{4m\beta^2}\left[\frac{1}{2}\left(1-e^{-\left(\frac{x-x_0}{\beta}\right)}\right)^2 -e^{-\left(\frac{x-x_0}{\beta}\right)}+\frac{3}{2} \right]
\end{equation}
and thus also find the energy associated with this wavefunction to be 
$E=3\hbar^2/(8 m \beta^2)$.  The potential $V_{\rm Gbl}$ 
is shown in Fig.~\ref{Fig:Gumbel:TISE} 
for select values of $x_0$ and $\beta$.  
The Gumbel potential $V_{\rm Gbl}(x)$ has a minimum at 
$x_{\rm min} = x_0-\beta\ln{(2)}$ which has $V(x_{\rm min})=0$.

%\FloatBarrier

\begin{figure}[htp]
    \centering
    \includegraphics[width=8.6 cm]{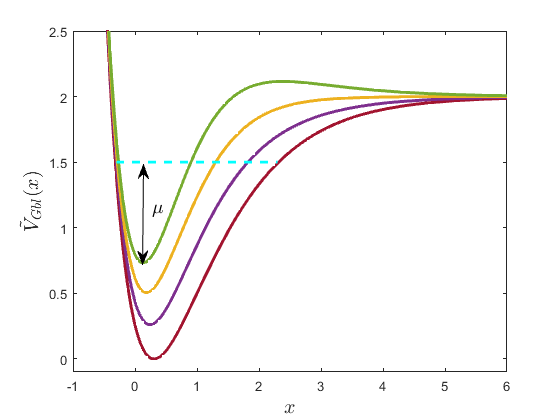}
    \caption{${\widetilde V}(x)$ for the 1D Gumbel distribution, with $\frac{\hbar^2}{4m}= 1$, $x_0 = 1$, and $\beta = 1$ for four different values of $gN$: $gN = 0$ (maroon), $gN = 1$ (magenta), $gN = 2$ (gold), and $gN = 3$ (green). 
    The chemical potential $\mu$ is the difference between the cyan dashed horizontal 
    line at $\frac{3}{2}=\frac{\hbar^2}{4m}\frac{3}{2\beta^2}$ and the minimum of the curve ${\widetilde V}(x)$, as shown for $gN=3$.  
    }
    \label{Fig:Gumbel:GPeq}
\end{figure}

For the 1D Gumbel distribution for the time-independent GPeq of Eq.~(\ref{Eq:GPeq}), 
the same procedure is followed.  
In fact, after solving for the pdf for the TISE, the \lq generalized' potential 
${\widetilde V}$ for the GPeq, Eq.~(\ref{Eq:VandV}), is found simply by adding the required 
nonlinear term to $V$ of the TISE.  
Explicitly for the 1D Gumbel pdf from Eq.~(\ref{Eq:VandV}) one obtains 
\begin{equation}
\label{Eq:Gumbel:VandV}
{\widetilde V}_{\rm Gbl}(x) 
= V_{\rm Gbl}(x) + g N P_{\rm Gbl}(x)
\end{equation}
with $V_{\rm Gbl}(x)$ from Eq.~(\ref{Eq:Gumbel:V}) and $P_{\rm Gbl}(x)$ from 
Eq.~(\ref{Eq:Gumbel:pdf}).  The chemical potential 
$\mu$ is the energy difference between $E=3\hbar^2/(8 m \beta^2)$ 
and the minimum of the generalized potential ${\widetilde V}(x)$.  
See Fig.~\ref{Fig:Gumbel:GPeq} for example plots of 
${\widetilde V}_{\rm Gbl}(x)$.  

%\FloatBarrier

We implemented our method to obtain solutions of the TISE for some well-known 1D pdfs with 
various domains.  
The exact potentials $V(x)$ and groundstate energy $E$ are provided in Table~I where 
column~3 contains the potentials for the pdfs in column~2.  

Four pdfs (Gaussian, Cauchy, Gumbel, and logistic) have potentials with zero walls 
since the domain is $x\in[-\infty,\infty]$. For the Gaussian potential, 
we can recognize that it is the simple harmonic oscillator  
(SHO or quadratic) potential with $V_{\rm min}$ at $x=x_0$ and $P(x)$ 
is the groundstate probability distribution. 
The Lorentzian (or Cauchy) potential has $V(x\longrightarrow\pm{\infty})\longrightarrow {\hbar^2}/{2m\gamma^2}$. 
This potential has a minimum at $x_{\rm min} = x_0 $ with $V_{(x_{min})} = 0$, 
and it has a maximum located at $x_{\rm max} = x_0 \pm{\gamma \sqrt{2}}$ with the value 
$V(x_{\rm max})= {2\hbar^2}/{3\gamma^2}$, but we have found one bound state for this potential.  
The logistic potential becomes zero at 
it's minimum $x=x_0$ and maximum at $x\longrightarrow \infty$. We found the groundstate 
energies (in units where $\hbar^2/4m=1$) to be $1/\sigma^2$, 
$2/\gamma^2$, and $1/{2s^2}$ for the Gaussian, Lorentzian, and 
logistic  distributions, respectively. 

We studied the two pdfs, Rayleigh and chi that have a potential with 1 wall since 
the domain is $x\in[0,\infty]$. The Rayleigh distribution is nothing but the chi 
distribution with two degrees of freedom, $k = 2$. The Rayleigh potential has its 
maximum at $x \pm \sigma$. We derived the potential which gives the generalization with $x\longrightarrow x/\sigma$ of the chi distribution with parameter $k$. 
Here we limit ourselves to the case where there is a finite minimum for 
$V(x)$ and a bound state, so the values in Table~I are only for $k\ge3$ which 
has the groundstate energy $E_0=\left(k-\sqrt{(k-3)(k-1)}\right)/\sigma^2$.

The 1D beta distribution has 2 hard walls 
since the domain is $x\in[0,1]$. For the beta potential, only
certain values of parameters $\alpha$ and $\beta$ give a potential $V(x)$ that does not go 
to $-\infty$. 
We outline four different cases for the beta potential 
considering possible values of the parameters $\alpha$ and $\beta$: 
1) $\lim_{x\to0^{+}} V(x) \longrightarrow  +\infty$ for $0<\alpha<1 \>{\rm or}\> 3<\alpha$; 
2) $\lim_{x\to0^{+}} V(x) \longrightarrow  -\infty$ for $1<\alpha<3$; 
3) $\lim_{x\to1^{-}} V(x) \longrightarrow +\infty$ for $0<\beta<1\ \>{\rm or}\> 3<\beta$ ;
and finally 
4) $\lim_{x \to 1^-} V(x) \longrightarrow -\infty$ for 
$ 1<\beta<3$.
Thus only in the case where both $0<\{\alpha,\beta\}<1$ or 
$3<\{\alpha,\beta\}$ is there a finite 
minimum for $V(x)$.  

The prescription outlined here for the TISE works for any 1D pdf, with modest 
sufficient constraints of being continuous and piece-wise twice 
continuously differentiable.  
Although the pdfs in Table~I all have a 
single maximum in $P(x)$, this is not a 
requirement and very complicated 
$P(x)$ can thought of that with the 
prescription gives very complicated $V(x)$.  
The prescription 
also works for a pdf that is an excited state, 
as is easily verified for 
example for the well-known excited state pdfs for the SHO potential.  
Solutions for Eq.~(\ref{Eq:TISE}) in higher dimensions are also 
easily found using this prescription.  In particular, for traditional 
separable pdfs, written in 2D Cartesian coordinates, 
$P(x,y)=P_x(x)P_y(y)$ gives 
$f(x,y)=f_x(x)+f_y(y)$ and hence a potential 
$V(x,y)=V_x(x)+V_y(y)$.  The same result generalizes 
to higher dimensions and to any other separable 
set of coordinates.  
One can argue the reason the prescription works so well is 
that it is the time-independent version of the 
nonlinear differential equation which corresponds to 
quantum mechanics as derived using Fisher information theory 
\cite{Frieden1989,Reginatto1998,Donker2016}, 
together with 
the {\it ansatz\/} $P({\vec r})=A\exp(-f({\vec r}))$.  
The equations derived using Fisher information are 
the analogue mathematically 
of Bohm's formulation of quantum mechanics
\cite{Bohm1952a,Bohm1952b}.  

Compare Eq.~(\ref{Eq:GPeq:f}) with 
Eq.~(\ref{Eq:TISE:f}), together with the 
definition of ${\widetilde V}({\vec r})$ of 
Eq.~(\ref{Eq:VandV}).  Thus in any case 
where the prescription works for the 
TISE there is a corresponding solution of the 
GPeq.  One illustration of this is 
in Fig.~\ref{Fig:Gumbel:GPeq} for the 
1D Gumbel distribution.  Of course one must 
keep in mind that at constant temperature 
and pressure for the Gibbs free energy $G$ 
that the chemical potential is 
$\mu=\left(\frac{\partial G}{\partial N}\right)_{T,P}$.

\begin{widetext}

\begin{table}[htb]
\caption{\label{tab:table3} 
Seven 1D pdfs, $P(x)$, that satisfy the TISE 
(so $g N = 0$) 
together with their 
potentials $V(x)$.  The mode of the top four pdfs is 
$x_0$.  
With the minimum of $V(x)$ set to the zero of energy, the 
energy of the groundstate is $E_0$.  
For the chi distribution the results are only for $k$$\ge$$3$.  
The beta distribution only has a minimum in $V(x)$ for certain 
values of $\alpha$ and $\beta$, as outlined in the text.  
}
\centering
\newpage
\setlength{\arrayrulewidth}{0.3mm}
\setlength{\tabcolsep}{17pt}
\renewcommand{\arraystretch}{3.0}
\resizebox{\textwidth}{!}{%
\centering
\begin{tabular}
{ |p{1.5cm}|p{3.0cm}|p{3.60cm}|p{1.2cm}|p{1.8cm}|  }
%\hline
%\multicolumn{5}{|c|}{\bf 1D Potentials} \\
\hline
$\quad$ pdf & $\qquad P\left(x\right)$ & 
$V\left(x\right)$ with 
$\left[\frac{\hbar^2}{4m} = 1\right]$
& $\>\>$ domain
& $E_0$$\quad$$\left[\frac{\hbar^2}{4m}=1\right]$   
\\
%\hline
\hline
Gaussian 
& $\frac{1}{\sigma\sqrt{2\pi}}e^{-\frac{1}{2}\left(\frac{x-x_0}{\sigma}\right)^2}$ 
& $\frac{(x-x_0)^2}{2\sigma^4}$  
& $-\infty$$<$$x$$<$$\infty$ & $\frac{1}{\sigma^2}$\\
%\hline %%%%%%%%%%%%%%%%%%%%
{\renewcommand{\arraystretch}{1.0}
$
\begin{array}{l}
{\rm Lorentzian} \\
\quad ({\rm Cauchy})
\end{array}
$ 
}
\renewcommand{\arraystretch}{1.6}
& $\frac{1}{\pi\gamma}\left[\frac{\gamma^2}{(x-x_0)^2+\gamma^2}\right]$   
& 
{\renewcommand{\arraystretch}{1.6}
$
\begin{array}{l}
\frac{6(x-x_0)^2}{\left[(x-x_0)^2+\gamma^2\right]^2} 
\\
\qquad - \>\> \frac{2}{(x-x_0)^2+\gamma^2}+\frac{2}{\gamma^2}  
\end{array} 
$
}
& $-\infty$$<$$x$$<$$\infty$
& $\frac{2}{\gamma^2} $ 
\\ 
%\hline %%%%%%%%%%%%%%%%%%%%%
Gumbel 
&
{\renewcommand{\arraystretch}{2.4}
$
\begin{array}{c} \frac{1}{\beta} \> 
\exp\Big[-\Big(\>\left(\frac{x-x_0}{\beta}\right)
\\
\>\>\> + \>\> e^{\left(- \frac{x-x_0}{\beta}\right)}\Big)\Big]
\end{array}$
}
& 
{\renewcommand{\arraystretch}{2.4}
$
\begin{array}{c}
\frac{1}{2\beta^2}\left(1-e^{-\left(\frac{x-x_0}{\beta}\right)}\right)^2 
\\
\qquad - \> 
\frac{e^{-\left(\frac{x-x_0}{\beta}\right)}}{\beta^2}+\frac{3}{2\beta^2}  
\end{array}
$
}
& $-\infty$$<$$x$$<$$\infty$
& $\frac{3}{2\beta^2}$  \\
%\hline %%%%%%%%%%%%%%%%%%%%
logistic  
& $\frac{1}{4s} \operatorname{sech^2}{\left(\frac{x-x_0}{2s}\right)}$ & $\frac{1}{s^2}\operatorname{tanh^2}\left(\frac{x-x_0}{2s}\right)$
& $-\infty$$<$$x$$<$$\infty$ & $\frac{1}{2s^2}$ \\
\hline
Rayleigh &$\frac{x}{\sigma^2}\exp{\left(-\frac{x^2}{2\sigma^2}\right)}$ &$\frac{1}{2}\left(\frac{x^2}{\sigma^4}-\frac{1}{x^2}\right)$
& $0$$\le$$x$$\le$$\infty$ & $\frac{2}{\sigma^2}$ \\
%\hline %%%%%%%%%%%%%%%%%%%%
chi & 
{\renewcommand{\arraystretch}{2.2}
$
\begin{array}{c}
\frac{1}{\sigma 2^{\frac{k-2}{2}}
\Gamma{\left(\frac{k}{2}\right)}}\left(\frac{x}{\sigma}\right)^{k-1}
\\ 
\>\>\> \times \> \exp\left(-\frac{x^2}{2\sigma^2}\right) 
\end{array}
$
}
& {\renewcommand{\arraystretch}{2.2}
$
\begin{array}{c}
\frac{x^2}{2\sigma^4}+\frac{(k-1)(k-3)}{2x^2}
\\
\qquad - \>\> \frac{\sqrt{(k-1)(k-3)}}{\sigma^2}
\end{array}
$
}
& $0$$\le$$x$$\le$$\infty$ & 
$\frac{k-\sqrt{(k-1)(k-3)}}{\sigma^2}$     
\\ 
\hline %%%%%%%%%%%%%%%%%%%%
beta 
& {\renewcommand{\arraystretch}{2.0}
$
\begin{array}{c}
\frac{\Gamma(\alpha+\beta)}{\Gamma(\alpha)\Gamma(\beta)}x^{\alpha -1}
\\ 
\qquad \times \>\> (1-x)^{\beta-1}
\end{array}
$
}
&  {\renewcommand{\arraystretch}{2.0}
$
\begin{array}{l}
\frac{(\alpha+\beta-2)(\alpha+\beta -4)x^2}{2x^2(1-x)^2}
\\
\qquad - \> \frac{2x(\alpha -1)(\alpha+\beta-4)}{2x^2(1-x)^2}
\\
\qquad +\> \frac{(\alpha^2-4\alpha+3)}{2x^2(1-x)^2} 
\end{array}
$
}
& $0$$\le$$x$$\le$$1$ & {\rm see text} \\ 
\hline %%%%%%%%%%%%%%%%%%%%
\end{tabular}}
\end{table}

\setlength{\arrayrulewidth}{0.4mm}
\setlength{\tabcolsep}{18pt}
\renewcommand{\arraystretch}{3.5}

To provide another concrete 2D example, 
we use the 2D pdf for 
the hydrogen atom in momentum space.  The H 
atom in momentum space has a long history 
starting in 1928 
\cite{podolsky1928quantum,podolsky1929momentum,lombardi1980hydrogen},
but we use the notation from the 2020 article 
\cite{lombardi2020hydrogen}.  
From \cite{lombardi2020hydrogen} we use their two 
variables 
for $p_r$ and $\theta_p$ and wavefunctions 
from their Eq.~[36], while performing the 
integral over their variable $\phi_p$.  
The unit for momentum we use is 
$p_0=2\pi \hbar/a_0$ with $a_0$ the Bohr radius of the 
hydrogen atom.  
These atomic physics pdfs 
$P_{n,\ell,m_\ell}(p_r,\theta_p)$ for the 
quantum numbers $n$, $\ell$, and $m_\ell$ for the 
hydrogen atom in momentum space can be viewed as 
just given examples of a pdf.  
We choose to plot both the pdf and the 
potential in Cartesian coordinates with 
$x'=p_r$ and $y'=\theta_p$.  
Consequently, we ask which 
$V_{n,\ell,m_\ell}(x',y')=V_{n,\ell,m_\ell}(p_r,\theta_p)$ 
gives the desired pdf $P_{n,\ell,m_\ell}(x',y')$.  
An example is shown in Fig.~\ref{Fig:Hatom210}.

\begin{figure}[htp]
    \centering
    \includegraphics[width=16.0 cm]{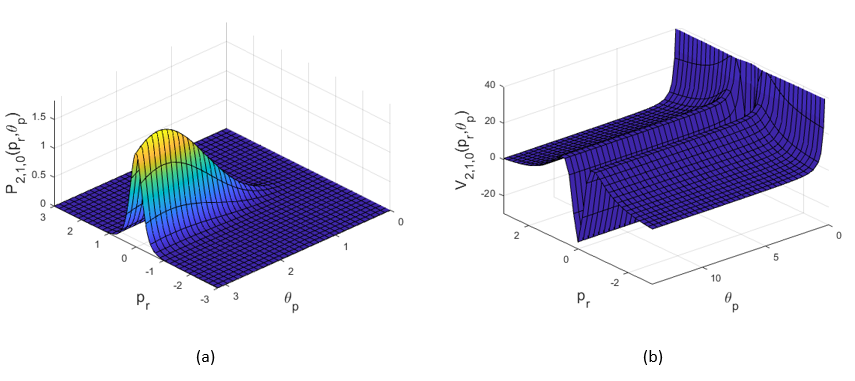}
    \caption{
    (a) The 2D pdf $P_{2,1,0}(p_r,\theta_p)$ for the 
    hydrogen atom in momentum space, plotted 
    in Cartesian coordinates $x'=p_r$ and $y'=\theta_p$.  
    The units used are $p_0=1$.  
    (b) The 2D potential $V_{2,1,0}(p_r,\theta_p)$ that 
    gives the 2D pdf $P_{2,1,0}(p_r,\theta_p)$.
    }
    \label{Fig:Hatom210}
\end{figure}
%\FloatBarrier

In summary, we have introduced a new 
methodology to solve 
both the time-independent 
Gross-Pitaevskii equation (GPeq) 
and the time-independent Schr{\"o}dinger 
equation (TISE).  The method of solving the 
TISE is simple enough to be taught to students 
the first time they are introduced to this 
differential equation.  The introduced 
prescription gives a wide class of 
new solutions for bound states of the TISE, 
and works in any dimension.  
The GPeq solutions obtained using the 
prescription are also relatively easy, and 
can be used for any value (and sign) of 
$g$.  For an atomic BEC the GPeq parameter 
$g$ can be of either sign as well as be 
large or small, and still 
the prescription will work.  
With experimentally obtained \lq painted' potentials 
\cite{ColdQuanta,Henderson2009} for an atomic 
BEC the prescription may have additional potential 
technological benefits.  The prescription will also 
work in other cases where the GPeq may be valid, 
for example in the case of $\psi$-dark-matter 
\cite{Hui2017}.  Future work will be to 
implement a related study for the 
time-dependent Schr{\"o}dinger equation 
and Gross-Pitaevskii equation for one or 
many {B}ose-{E}instein condensates 
\cite{Egorov2013}.  

\end{widetext}

\clearpage

\begin{acknowledgements}
MAN thanks H.~{D}e~{R}aedt and K.~Michielsen for 
useful early discussions related to 
Eq.~(\ref{Eq:define:f}) and Fisher information 
in quantum mechanics.  
\end{acknowledgements}

\bibliography{TISEGP}% Produces the bibliography via BibTeX.

\providecommand{\noopsort}[1]{}\providecommand{\singleletter}[1]{#1}%
\begin{thebibliography}{31}%
\makeatletter
\providecommand \@ifxundefined [1]{%
 \@ifx{#1\undefined}
}%
\providecommand \@ifnum [1]{%
 \ifnum #1\expandafter \@firstoftwo
 \else \expandafter \@secondoftwo
 \fi
}%
\providecommand \@ifx [1]{%
 \ifx #1\expandafter \@firstoftwo
 \else \expandafter \@secondoftwo
 \fi
}%
\providecommand \natexlab [1]{#1}%
\providecommand \enquote  [1]{``#1''}%
\providecommand \bibnamefont  [1]{#1}%
\providecommand \bibfnamefont [1]{#1}%
\providecommand \citenamefont [1]{#1}%
\providecommand \href@noop [0]{\@secondoftwo}%
\providecommand \href [0]{\begingroup \@sanitize@url \@href}%
\providecommand \@href[1]{\@@startlink{#1}\@@href}%
\providecommand \@@href[1]{\endgroup#1\@@endlink}%
\providecommand \@sanitize@url [0]{\catcode `\\12\catcode `\$12\catcode
  `\&12\catcode `\#12\catcode `\^12\catcode `\_12\catcode `\%12\relax}%
\providecommand \@@startlink[1]{}%
\providecommand \@@endlink[0]{}%
\providecommand \url  [0]{\begingroup\@sanitize@url \@url }%
\providecommand \@url [1]{\endgroup\@href {#1}{\urlprefix }}%
\providecommand \urlprefix  [0]{URL }%
\providecommand \Eprint [0]{\href }%
\providecommand \doibase [0]{https://doi.org/}%
\providecommand \selectlanguage [0]{\@gobble}%
\providecommand \bibinfo  [0]{\@secondoftwo}%
\providecommand \bibfield  [0]{\@secondoftwo}%
\providecommand \translation [1]{[#1]}%
\providecommand \BibitemOpen [0]{}%
\providecommand \bibitemStop [0]{}%
\providecommand \bibitemNoStop [0]{.\EOS\space}%
\providecommand \EOS [0]{\spacefactor3000\relax}%
\providecommand \BibitemShut  [1]{\csname bibitem#1\endcsname}%
\let\auto@bib@innerbib\@empty
%</preamble>
\bibitem [{\citenamefont {Bose}(1924)}]{bose1924plancks}%
  \BibitemOpen
  \bibfield  {author} {\bibinfo {author} {\bibfnamefont {S.~N.}\ \bibnamefont
  {Bose}},\ }\bibfield  {title} {\bibinfo {title} {Plancks gesetz und
  lichtquantenhypothese},\ }\href@noop {} {\bibfield  {journal} {\bibinfo
  {journal} {Zeitschrift für Physik}\ }\textbf {\bibinfo {volume} {26}},\
  \bibinfo {pages} {178} (\bibinfo {year} {1924})}\BibitemShut {NoStop}%
\bibitem [{\citenamefont {Einstein}(1925)}]{einstein2005quantentheorie}%
  \BibitemOpen
  \bibfield  {author} {\bibinfo {author} {\bibfnamefont {A.}~\bibnamefont
  {Einstein}},\ }\bibfield  {title} {\bibinfo {title} {Quantentheorie des
  einatomigen idealen {G}ases},\ }\href@noop {} {\bibfield  {journal} {\bibinfo
   {journal} {Akademie-Vortr{\"a}ge: Sitzungsberichte der Preu{\ss}ischen
  Akademie der Wissenschaften, Sitzung der physikalish-mathematisehen}\ ,\
  \bibinfo {pages} {(12~pages)}} (\bibinfo {year} {1925})}\BibitemShut
  {NoStop}%
\bibitem [{\citenamefont {Petrich}\ \emph {et~al.}(1995)\citenamefont
  {Petrich}, \citenamefont {Anderson}, \citenamefont {J.R.},\ and\
  \citenamefont {Cornell}}]{Petrich1995}%
  \BibitemOpen
  \bibfield  {author} {\bibinfo {author} {\bibfnamefont {W.}~\bibnamefont
  {Petrich}}, \bibinfo {author} {\bibfnamefont {M.}~\bibnamefont {Anderson}},
  \bibinfo {author} {\bibfnamefont {E.}~\bibnamefont {J.R.}},\ and\ \bibinfo
  {author} {\bibfnamefont {E.}~\bibnamefont {Cornell}},\ }\bibfield  {title}
  {\bibinfo {title} {Stable, tightly confining magnetic trap for evaporative
  cooling of neutral atoms},\ }\href@noop {} {\bibfield  {journal} {\bibinfo
  {journal} {Phys. Rev. Lett.}\ }\textbf {\bibinfo {volume} {74}},\ \bibinfo
  {pages} {3352} (\bibinfo {year} {1995})}\BibitemShut {NoStop}%
\bibitem [{\citenamefont {Anderson}\ \emph {et~al.}(1995)\citenamefont
  {Anderson}, \citenamefont {Ensher}, \citenamefont {Matthews}, \citenamefont
  {C.E.},\ and\ \citenamefont {Cornell}}]{Anderson1995}%
  \BibitemOpen
  \bibfield  {author} {\bibinfo {author} {\bibfnamefont {M.}~\bibnamefont
  {Anderson}}, \bibinfo {author} {\bibfnamefont {J.}~\bibnamefont {Ensher}},
  \bibinfo {author} {\bibfnamefont {M.}~\bibnamefont {Matthews}}, \bibinfo
  {author} {\bibfnamefont {W.}~\bibnamefont {C.E.}},\ and\ \bibinfo {author}
  {\bibfnamefont {E.}~\bibnamefont {Cornell}},\ }\bibfield  {title} {\bibinfo
  {title} {Observation of {B}ose={E}instein condensation in a dilute atomic
  vapor},\ }\href@noop {} {\bibfield  {journal} {\bibinfo  {journal} {Science}\
  }\textbf {\bibinfo {volume} {269}},\ \bibinfo {pages} {198} (\bibinfo {year}
  {1995})}\BibitemShut {NoStop}%
\bibitem [{\citenamefont {Davis}\ \emph {et~al.}(1995)\citenamefont {Davis},
  \citenamefont {Mewes}, \citenamefont {Andrews}, \citenamefont {{v}an
  {D}ruten}, \citenamefont {Durfee}, \citenamefont {Kurn},\ and\ \citenamefont
  {Ketterle}}]{davis1995}%
  \BibitemOpen
  \bibfield  {author} {\bibinfo {author} {\bibfnamefont {K.}~\bibnamefont
  {Davis}}, \bibinfo {author} {\bibfnamefont {M.-O.}\ \bibnamefont {Mewes}},
  \bibinfo {author} {\bibfnamefont {M.}~\bibnamefont {Andrews}}, \bibinfo
  {author} {\bibfnamefont {N.}~\bibnamefont {{v}an {D}ruten}}, \bibinfo
  {author} {\bibfnamefont {D.}~\bibnamefont {Durfee}}, \bibinfo {author}
  {\bibfnamefont {D.}~\bibnamefont {Kurn}},\ and\ \bibinfo {author}
  {\bibfnamefont {W.}~\bibnamefont {Ketterle}},\ }\bibfield  {title} {\bibinfo
  {title} {Bose-{E}instein condensation in a gas of sodium atoms},\ }\href@noop
  {} {\bibfield  {journal} {\bibinfo  {journal} {Phys. Rev. Lett.}\ }\textbf
  {\bibinfo {volume} {75}},\ \bibinfo {pages} {3969} (\bibinfo {year}
  {1995})}\BibitemShut {NoStop}%
\bibitem [{\citenamefont {Cornell}\ and\ \citenamefont
  {Wieman}(2002)}]{cornell2002nobel}%
  \BibitemOpen
  \bibfield  {author} {\bibinfo {author} {\bibfnamefont {E.~A.}\ \bibnamefont
  {Cornell}}\ and\ \bibinfo {author} {\bibfnamefont {C.~E.}\ \bibnamefont
  {Wieman}},\ }\bibfield  {title} {\bibinfo {title} {Nobel lecture:
  {B}ose-{E}instein condensation in a dilute gas, the first 70 years and some
  recent experiments},\ }\href@noop {} {\bibfield  {journal} {\bibinfo
  {journal} {Rev. Modern Physics}\ }\textbf {\bibinfo {volume} {74}},\ \bibinfo
  {pages} {875} (\bibinfo {year} {2002})}\BibitemShut {NoStop}%
\bibitem [{\citenamefont {Ketterle}(2002)}]{ketterle2002nobel}%
  \BibitemOpen
  \bibfield  {author} {\bibinfo {author} {\bibfnamefont {W.}~\bibnamefont
  {Ketterle}},\ }\bibfield  {title} {\bibinfo {title} {Nobel lecture: When
  atoms behave as waves: {B}ose-{E}instein condensation and the atom laser},\
  }\href@noop {} {\bibfield  {journal} {\bibinfo  {journal} {Rev. Modern
  Physics}\ }\textbf {\bibinfo {volume} {74}},\ \bibinfo {pages} {1131}
  (\bibinfo {year} {2002})}\BibitemShut {NoStop}%
\bibitem [{\citenamefont {Amico}\ \emph {et~al.}(2021)\citenamefont {Amico}
  \emph {et~al.}}]{amico2021roadmap}%
  \BibitemOpen
  \bibfield  {author} {\bibinfo {author} {\bibfnamefont {L.}~\bibnamefont
  {Amico}} \emph {et~al.},\ }\bibfield  {title} {\bibinfo {title} {Roadmap on
  atomtronics: State of the art and perspective},\ }\href@noop {} {\bibfield
  {journal} {\bibinfo  {journal} {AVS Quantum Science}\ }\textbf {\bibinfo
  {volume} {3}},\ \bibinfo {pages} {039201} (\bibinfo {year}
  {2021})}\BibitemShut {NoStop}%
\bibitem [{\citenamefont {Pepino}(2021)}]{pepino2021advances}%
  \BibitemOpen
  \bibfield  {author} {\bibinfo {author} {\bibfnamefont {R.~A.}\ \bibnamefont
  {Pepino}},\ }\bibfield  {title} {\bibinfo {title} {Advances in atomtronics},\
  }\href@noop {} {\bibfield  {journal} {\bibinfo  {journal} {Entropy}\ }\textbf
  {\bibinfo {volume} {23}},\ \bibinfo {pages} {534} (\bibinfo {year}
  {2021})}\BibitemShut {NoStop}%
\bibitem [{Col(2007)}]{ColdQuanta}%
  \BibitemOpen
  \href@noop {} {\bibinfo {title} {{ColdQuanta Albert}}},\ \bibinfo
  {howpublished} {\url{https://bec.coldquantaapis.com/about}} (\bibinfo {year}
  {2007})\BibitemShut {NoStop}%
\bibitem [{\citenamefont {Henderson}\ \emph {et~al.}(2009)\citenamefont
  {Henderson}, \citenamefont {Ryu}, \citenamefont {Mac{C}ormick},\ and\
  \citenamefont {Boshier}}]{Henderson2009}%
  \BibitemOpen
  \bibfield  {author} {\bibinfo {author} {\bibfnamefont {K.}~\bibnamefont
  {Henderson}}, \bibinfo {author} {\bibfnamefont {C.}~\bibnamefont {Ryu}},
  \bibinfo {author} {\bibfnamefont {C.}~\bibnamefont {Mac{C}ormick}},\ and\
  \bibinfo {author} {\bibfnamefont {M.}~\bibnamefont {Boshier}},\ }\bibfield
  {title} {\bibinfo {title} {Experimental demonstration of painting arbitrary
  and dynamic potentials for {B}ose–{E}instein condensates},\ }\href@noop {}
  {\bibfield  {journal} {\bibinfo  {journal} {New J.\ Phys.}\ }\textbf
  {\bibinfo {volume} {11}},\ \bibinfo {pages} {043020} (\bibinfo {year}
  {2009})}\BibitemShut {NoStop}%
\bibitem [{\citenamefont {Erd{\"o}s}\ \emph {et~al.}(2007)\citenamefont
  {Erd{\"o}s}, \citenamefont {Schlein},\ and\ \citenamefont {Yau}}]{ERDOS2007}%
  \BibitemOpen
  \bibfield  {author} {\bibinfo {author} {\bibfnamefont {L.}~\bibnamefont
  {Erd{\"o}s}}, \bibinfo {author} {\bibfnamefont {B.}~\bibnamefont {Schlein}},\
  and\ \bibinfo {author} {\bibfnamefont {H.-T.}\ \bibnamefont {Yau}},\
  }\bibfield  {title} {\bibinfo {title} {Rigorous derivation of the
  {G}ross-{P}itaevskii equation},\ }\href@noop {} {\bibfield  {journal}
  {\bibinfo  {journal} {Phys. {R}ev.\ {L}ett.}\ }\textbf {\bibinfo {volume}
  {98}},\ \bibinfo {pages} {040404} (\bibinfo {year} {2007})}\BibitemShut
  {NoStop}%
\bibitem [{\citenamefont {Gross}(1961)}]{gross1961structure}%
  \BibitemOpen
  \bibfield  {author} {\bibinfo {author} {\bibfnamefont {E.~P.}\ \bibnamefont
  {Gross}},\ }\bibfield  {title} {\bibinfo {title} {Structure of a quantized
  vortex in boson systems},\ }\href@noop {} {\bibfield  {journal} {\bibinfo
  {journal} {Il Nuovo Cimento (1955-1965)}\ }\textbf {\bibinfo {volume} {20}},\
  \bibinfo {pages} {454} (\bibinfo {year} {1961})}\BibitemShut {NoStop}%
\bibitem [{\citenamefont {Pitaevskii}(1961)}]{pitaevskii1961vortex}%
  \BibitemOpen
  \bibfield  {author} {\bibinfo {author} {\bibfnamefont {L.~P.}\ \bibnamefont
  {Pitaevskii}},\ }\bibfield  {title} {\bibinfo {title} {Vortex lines in an
  imperfect {B}ose gas},\ }\href@noop {} {\bibfield  {journal} {\bibinfo
  {journal} {Sov. Phys. JETP}\ }\textbf {\bibinfo {volume} {13}},\ \bibinfo
  {pages} {451} (\bibinfo {year} {1961})}\BibitemShut {NoStop}%
\bibitem [{\citenamefont {Meyer}\ and\ \citenamefont
  {Wong}(2014)}]{meyer2014quantum}%
  \BibitemOpen
  \bibfield  {author} {\bibinfo {author} {\bibfnamefont {D.~A.}\ \bibnamefont
  {Meyer}}\ and\ \bibinfo {author} {\bibfnamefont {T.~G.}\ \bibnamefont
  {Wong}},\ }\bibfield  {title} {\bibinfo {title} {Quantum search with general
  nonlinearities},\ }\href@noop {} {\bibfield  {journal} {\bibinfo  {journal}
  {Physical Review A}\ }\textbf {\bibinfo {volume} {89}},\ \bibinfo {pages}
  {012312} (\bibinfo {year} {2014})}\BibitemShut {NoStop}%
\bibitem [{\citenamefont {Donley}\ \emph {et~al.}(2001)\citenamefont {Donley},
  \citenamefont {Claussen}, \citenamefont {Cornish}, \citenamefont {Roberts},
  \citenamefont {Cornell},\ and\ \citenamefont {Wieman}}]{Donley2001}%
  \BibitemOpen
  \bibfield  {author} {\bibinfo {author} {\bibfnamefont {E.}~\bibnamefont
  {Donley}}, \bibinfo {author} {\bibfnamefont {N.}~\bibnamefont {Claussen}},
  \bibinfo {author} {\bibfnamefont {S.}~\bibnamefont {Cornish}}, \bibinfo
  {author} {\bibfnamefont {J.}~\bibnamefont {Roberts}}, \bibinfo {author}
  {\bibfnamefont {E.}~\bibnamefont {Cornell}},\ and\ \bibinfo {author}
  {\bibfnamefont {C.}~\bibnamefont {Wieman}},\ }\bibfield  {title} {\bibinfo
  {title} {Dynamics of collapsing and exploding {B}ose-{E}instein
  condensates},\ }\href@noop {} {\bibfield  {journal} {\bibinfo  {journal}
  {Nature}\ }\textbf {\bibinfo {volume} {412}},\ \bibinfo {pages} {296}
  (\bibinfo {year} {2001})}\BibitemShut {NoStop}%
\bibitem [{\citenamefont {Altin}\ \emph {et~al.}(2010)\citenamefont {Altin},
  \citenamefont {Robins}, \citenamefont {D{\"o}ring}, \citenamefont {Debs},
  \citenamefont {Poldy}, \citenamefont {Figl},\ and\ \citenamefont
  {Close}}]{Altin2010}%
  \BibitemOpen
  \bibfield  {author} {\bibinfo {author} {\bibfnamefont {P.}~\bibnamefont
  {Altin}}, \bibinfo {author} {\bibfnamefont {N.}~\bibnamefont {Robins}},
  \bibinfo {author} {\bibfnamefont {D.}~\bibnamefont {D{\"o}ring}}, \bibinfo
  {author} {\bibfnamefont {J.}~\bibnamefont {Debs}}, \bibinfo {author}
  {\bibfnamefont {R.}~\bibnamefont {Poldy}}, \bibinfo {author} {\bibfnamefont
  {C.}~\bibnamefont {Figl}},\ and\ \bibinfo {author} {\bibfnamefont
  {J.}~\bibnamefont {Close}},\ }\bibfield  {title} {\bibinfo {title}
  {$^{85}${R}b tunable-interaction {Bose}-{E}instein condensate machine},\
  }\href@noop {} {\bibfield  {journal} {\bibinfo  {journal} {Rev. Scientific
  Instruments}\ }\textbf {\bibinfo {volume} {81}},\ \bibinfo {pages} {063103}
  (\bibinfo {year} {2010})}\BibitemShut {NoStop}%
\bibitem [{\citenamefont {Davis}(2001)}]{davis2001dynamics}%
  \BibitemOpen
  \bibfield  {author} {\bibinfo {author} {\bibfnamefont {M.~J.}\ \bibnamefont
  {Davis}},\ }\emph {\bibinfo {title} {Dynamics of {B}ose-{E}instein
  condensation}},\ \href@noop {} {Ph.D. thesis},\ \bibinfo  {school} {Citeseer}
  (\bibinfo {year} {2001})\BibitemShut {NoStop}%
\bibitem [{\citenamefont {Griffiths}\ and\ \citenamefont
  {Schroeter}(2018)}]{griffiths2018introduction}%
  \BibitemOpen
  \bibfield  {author} {\bibinfo {author} {\bibfnamefont {D.~J.}\ \bibnamefont
  {Griffiths}}\ and\ \bibinfo {author} {\bibfnamefont {D.~F.}\ \bibnamefont
  {Schroeter}},\ }\href@noop {} {\emph {\bibinfo {title} {Introduction to
  {Q}uantum {M}echanics}}}\ (\bibinfo  {publisher} {Cambridge University
  Press},\ \bibinfo {year} {2018})\BibitemShut {NoStop}%
\bibitem [{\citenamefont {McIntyre}\ \emph {et~al.}(2012)\citenamefont
  {McIntyre}, \citenamefont {Manogue},\ and\ \citenamefont
  {Tate}}]{mcintyre2012quantum}%
  \BibitemOpen
  \bibfield  {author} {\bibinfo {author} {\bibfnamefont {D.~H.}\ \bibnamefont
  {McIntyre}}, \bibinfo {author} {\bibfnamefont {C.~A.}\ \bibnamefont
  {Manogue}},\ and\ \bibinfo {author} {\bibfnamefont {J.}~\bibnamefont
  {Tate}},\ }\href@noop {} {\emph {\bibinfo {title} {Quantum {M}echanics: {A}
  paradigms approach}}},\ Vol.\ \bibinfo {volume} {192}\ (\bibinfo  {publisher}
  {Pearson Boston},\ \bibinfo {year} {2012})\BibitemShut {NoStop}%
\bibitem [{\citenamefont {Frieden}(1989)}]{Frieden1989}%
  \BibitemOpen
  \bibfield  {author} {\bibinfo {author} {\bibfnamefont {B.}~\bibnamefont
  {Frieden}},\ }\bibfield  {title} {\bibinfo {title} {Fisher information as the
  basis for the {S}chr{\"o}dinger wave equation},\ }\href@noop {} {\bibfield
  {journal} {\bibinfo  {journal} {Am. J. Phys.}\ }\textbf {\bibinfo {volume}
  {57}},\ \bibinfo {pages} {1004} (\bibinfo {year} {1989})}\BibitemShut
  {NoStop}%
\bibitem [{\citenamefont {Reginatto}(1998)}]{Reginatto1998}%
  \BibitemOpen
  \bibfield  {author} {\bibinfo {author} {\bibfnamefont {M.}~\bibnamefont
  {Reginatto}},\ }\bibfield  {title} {\bibinfo {title} {Derivation of the
  equations of nonrelativistic quantum mechanics using the principle of minimum
  {F}isher information},\ }\href@noop {} {\bibfield  {journal} {\bibinfo
  {journal} {Phys. Rev. A}\ }\textbf {\bibinfo {volume} {58}},\ \bibinfo
  {pages} {1775–1778} (\bibinfo {year} {1998})}\BibitemShut {NoStop}%
\bibitem [{\citenamefont {Donker}\ \emph {et~al.}(2016)\citenamefont {Donker},
  \citenamefont {Katsnelson}, \citenamefont {{D}e {R}aedt},\ and\ \citenamefont
  {Michielsen}}]{Donker2016}%
  \BibitemOpen
  \bibfield  {author} {\bibinfo {author} {\bibfnamefont {H.}~\bibnamefont
  {Donker}}, \bibinfo {author} {\bibfnamefont {M.}~\bibnamefont {Katsnelson}},
  \bibinfo {author} {\bibfnamefont {H.}~\bibnamefont {{D}e {R}aedt}},\ and\
  \bibinfo {author} {\bibfnamefont {K.}~\bibnamefont {Michielsen}},\ }\bibfield
   {title} {\bibinfo {title} {Logical inference approach to relativistic
  quantum mechanics: {D}erivation of the {K}lein-{G}ordon equation},\
  }\href@noop {} {\bibfield  {journal} {\bibinfo  {journal} {Annals Phys.}\
  }\textbf {\bibinfo {volume} {372}},\ \bibinfo {pages} {74} (\bibinfo {year}
  {2016})}\BibitemShut {NoStop}%
\bibitem [{\citenamefont {Bohm}(1952{\natexlab{a}})}]{Bohm1952a}%
  \BibitemOpen
  \bibfield  {author} {\bibinfo {author} {\bibfnamefont {D.}~\bibnamefont
  {Bohm}},\ }\bibfield  {title} {\bibinfo {title} {A suggested interpretation
  of the quantum theory in terms of \lq\lq hidden'' variables. {I}},\
  }\href@noop {} {\bibfield  {journal} {\bibinfo  {journal} {Phys. Rev.}\
  }\textbf {\bibinfo {volume} {85}},\ \bibinfo {pages} {166} (\bibinfo {year}
  {1952}{\natexlab{a}})}\BibitemShut {NoStop}%
\bibitem [{\citenamefont {Bohm}(1952{\natexlab{b}})}]{Bohm1952b}%
  \BibitemOpen
  \bibfield  {author} {\bibinfo {author} {\bibfnamefont {D.}~\bibnamefont
  {Bohm}},\ }\bibfield  {title} {\bibinfo {title} {A suggested interpretation
  of the quantum theory in terms of \lq\lq hidden'' variables. {II}},\
  }\href@noop {} {\bibfield  {journal} {\bibinfo  {journal} {Phys. Rev.}\
  }\textbf {\bibinfo {volume} {85}},\ \bibinfo {pages} {180} (\bibinfo {year}
  {1952}{\natexlab{b}})}\BibitemShut {NoStop}%
\bibitem [{\citenamefont {Podolsky}(1928)}]{podolsky1928quantum}%
  \BibitemOpen
  \bibfield  {author} {\bibinfo {author} {\bibfnamefont {B.}~\bibnamefont
  {Podolsky}},\ }\bibfield  {title} {\bibinfo {title} {Quantum-mechanically
  correct form of {H}amiltonian function for conservative systems},\
  }\href@noop {} {\bibfield  {journal} {\bibinfo  {journal} {Physical Review}\
  }\textbf {\bibinfo {volume} {32}},\ \bibinfo {pages} {812} (\bibinfo {year}
  {1928})}\BibitemShut {NoStop}%
\bibitem [{\citenamefont {Podolsky}\ and\ \citenamefont
  {Pauling}(1929)}]{podolsky1929momentum}%
  \BibitemOpen
  \bibfield  {author} {\bibinfo {author} {\bibfnamefont {B.}~\bibnamefont
  {Podolsky}}\ and\ \bibinfo {author} {\bibfnamefont {L.}~\bibnamefont
  {Pauling}},\ }\bibfield  {title} {\bibinfo {title} {The momentum distribution
  in hydrogen-like atoms},\ }\href@noop {} {\bibfield  {journal} {\bibinfo
  {journal} {Physical Review}\ }\textbf {\bibinfo {volume} {34}},\ \bibinfo
  {pages} {109} (\bibinfo {year} {1929})}\BibitemShut {NoStop}%
\bibitem [{\citenamefont {Lombardi}(1980)}]{lombardi1980hydrogen}%
  \BibitemOpen
  \bibfield  {author} {\bibinfo {author} {\bibfnamefont {J.~R.}\ \bibnamefont
  {Lombardi}},\ }\bibfield  {title} {\bibinfo {title} {Hydrogen atom in the
  momentum representation},\ }\href@noop {} {\bibfield  {journal} {\bibinfo
  {journal} {Physical Review A}\ }\textbf {\bibinfo {volume} {22}},\ \bibinfo
  {pages} {797} (\bibinfo {year} {1980})}\BibitemShut {NoStop}%
\bibitem [{\citenamefont {Lombardi}\ and\ \citenamefont
  {Ogilvie}(2020)}]{lombardi2020hydrogen}%
  \BibitemOpen
  \bibfield  {author} {\bibinfo {author} {\bibfnamefont {J.}~\bibnamefont
  {Lombardi}}\ and\ \bibinfo {author} {\bibfnamefont {J.}~\bibnamefont
  {Ogilvie}},\ }\bibfield  {title} {\bibinfo {title} {The hydrogen atom in the
  momentum representation; a critique of the variables comprising the momentum
  representation},\ }\href@noop {} {\bibfield  {journal} {\bibinfo  {journal}
  {Chemical Physics}\ }\textbf {\bibinfo {volume} {538}},\ \bibinfo {pages}
  {110886} (\bibinfo {year} {2020})}\BibitemShut {NoStop}%
\bibitem [{\citenamefont {Hui}\ \emph {et~al.}(2017)\citenamefont {Hui},
  \citenamefont {Ostriker}, \citenamefont {Tremaine},\ and\ \citenamefont
  {Witten}}]{Hui2017}%
  \BibitemOpen
  \bibfield  {author} {\bibinfo {author} {\bibfnamefont {L.}~\bibnamefont
  {Hui}}, \bibinfo {author} {\bibfnamefont {J.}~\bibnamefont {Ostriker}},
  \bibinfo {author} {\bibfnamefont {S.}~\bibnamefont {Tremaine}},\ and\
  \bibinfo {author} {\bibfnamefont {E.}~\bibnamefont {Witten}},\ }\bibfield
  {title} {\bibinfo {title} {Ultralight scalars as cosmological dark matter},\
  }\href@noop {} {\bibfield  {journal} {\bibinfo  {journal} {Phys. Rev D}\
  }\textbf {\bibinfo {volume} {95}},\ \bibinfo {pages} {043541} (\bibinfo
  {year} {2017})}\BibitemShut {NoStop}%
\bibitem [{\citenamefont {Egorov}\ \emph {et~al.}(2013)\citenamefont {Egorov},
  \citenamefont {Opanchuk}, \citenamefont {Drummond}, \citenamefont {Hall},
  \citenamefont {Hannaford},\ and\ \citenamefont {Sidorov}}]{Egorov2013}%
  \BibitemOpen
  \bibfield  {author} {\bibinfo {author} {\bibfnamefont {M.}~\bibnamefont
  {Egorov}}, \bibinfo {author} {\bibfnamefont {B.}~\bibnamefont {Opanchuk}},
  \bibinfo {author} {\bibfnamefont {P.}~\bibnamefont {Drummond}}, \bibinfo
  {author} {\bibfnamefont {B.}~\bibnamefont {Hall}}, \bibinfo {author}
  {\bibfnamefont {P.}~\bibnamefont {Hannaford}},\ and\ \bibinfo {author}
  {\bibfnamefont {A.}~\bibnamefont {Sidorov}},\ }\bibfield  {title} {\bibinfo
  {title} {Measurement of $s$-wave scattering lengths in a two-component
  {B}ose-{E}instein condensate},\ }\href@noop {} {\bibfield  {journal}
  {\bibinfo  {journal} {Phys. Rev. A}\ }\textbf {\bibinfo {volume} {87}},\
  \bibinfo {pages} {053614} (\bibinfo {year} {2013})}\BibitemShut {NoStop}%
\end{thebibliography}%

\end{document}